\documentclass{elsart}

 \usepackage{graphics}
 \usepackage{epsfig}

\usepackage{amssymb}

\begin{document}

\begin{frontmatter}

  \title{Flux profile scanners for scattered high--energy electrons}
  \author[UMass]{R.S. Hicks\thanksref{Ross}},
  \author[Smith]{P. Decowski\thanksref{Piotr}},
  \author[UMass]{C. Arroyo},
  \author[UMass]{M. Breuer\thanksref{Mat}},
  \author[UMass]{J. Celli\thanksref{Celli}},
  \author[JLab]{E. Chudakov},
  \author[UMass]{K.S. Kumar},
  \author[SLAC]{M. Olson\thanksref{MO}},
  \author[UMass]{G.A. Peterson},
  \author[Smith]{K. Pope\thanksref{Kat}},
  \author[UMass]{J. Ricci},
  \author[UMass]{J. Savage},
  \author[Syr]{and P.A. Souder}
  \thanks[Ross]{Present address: 5037 Great Ocean Rd., Lavers Hill, Victoria, Australia 3238.}
  \thanks[Piotr]{Corresponding author. Department of Physics, Smith College, 
Northampton, MA, 01063, USA, Tel: 413 585 3882, FAX 413 585 3786, E--mail: pdecowski@smith.edu.}
  \thanks[Mat]{Present address: 97 Mt. Warner Rd., Hadley, MA 01035, USA.}
  \thanks[Celli]{Present address: Physics Dept., 590 Commonwealth Ave., 
Boston University, Boston, MA 02215, USA.}
  \thanks[MO]{Present address: St. Norbert College, De Pere, WI 54115, USA.}
  \thanks[Kat]{Deceased.}
  \address[UMass]{University of Massachusetts, Amherst, MA 01003, USA}
  \address[Smith]{Smith College, Northampton, MA 01063, USA}
  \address[JLab]{Thomas Jefferson Laboratory, Newport News, VA 23606, USA}
  \address[SLAC]{Stanford Linear Accelerator Center, Menlo Park, CA 94025, USA}
  \address[Syr]{University of Syracuse, Syracuse, NY 13244, USA}

\begin{abstract}
The paper describes the design and performance of flux integrating
Cherenkov scanners with air--core reflecting light guides used in a high--energy, high--flux
electron scattering experiment at the Stanford Linear Accelerator Center.  The scanners
were highly radiation resistant and provided a good signal to background ratio leading to
very good spatial resolution of the scattered electron flux profile scans.
\end{abstract}

\begin{keyword}
 Electron flux profile scanner \sep Cherenkov detector \sep air--core reflecting light
guide.

\PACS 
29.40.Ka
\end{keyword}
\end{frontmatter}

\section{Introduction}
\label{intro}
 Electron--electron (M{\o}ller) scattering at GeV beam energies gives rise to a far--forward
angular  distribution  of  scattered  electrons.   If  hydrogen  targets  are  used,  small--angle
electron  scattering  from  protons  must  be  distinguished  from  the  signal of interest which is 
the electron--electron
scattering.   This report describes detectors built to measure the spatial distribution and
relative integral flux of scattered electrons from intense incident beams of 48.3 and 45.0
GeV  electrons  incident  upon  a  liquid  hydrogen  target  in  the  E158  experiment  at  the
Stanford Linear Accelerator Center \cite{e158}.
  
The detector design confronted two main challenges. First, the intensity of the scattered
electron flux of interest ($ \sim 10^{11}$ electrons/s/cm$^2$ at a duty cycle of $2.4*10^{-5}$) 
created a very harsh radiation
environment. This precluded the use of components that would be exposed for significant
periods  to  the  scattered  beam,  including  photomultiplier  tubes.  Second,  within  the
scattered beam the particle flux varied by more than an order--of--magnitude.  The detector
therefore needed to have fine granularity without being influenced by the magnitude of
the flux nearby.

   The  active  detector  element  was  fused silica that produced  Cherenkov  light  and  was
preceded by a tungsten pre--radiator.  Such detectors are not only radiation--hard, but they
also have a directional response that can be exploited to suppress background.  Cherenkov
light from the fused silica was directed by means of a reflective air--core light--guide to a
photomultiplier  tube  (PMT)  safely  located  outside  the  scattered  beam,  50  cm  from  the
radiator.  Conventional solid light--guide materials were unacceptable.  Such light guides
would produce abundant Cherenkov light on their own and could scintillate.  Furthermore, they would
degrade  or  discolor  because  of  the  intense  radiation  background.   The  light  produced
inside  such  light  guides  could  exceed  the  small  signal  produced  when  the  detector's
active element $-$ but not the light guide $-$ is located in a region of relatively low scattered
beam intensity.   The air--core light guides we developed consisted of hollow--tubes lined
with  reflective  Alzak  sheet  \cite{Alzak}.   Alzak  has  an  optical  finish  produced  by
electrochemically brightening and anodizing a high purity aluminum alloy.   It also has
high permanent reflectivity and good resistance to corrosion and abrasion.

     These detectors were mounted on electromechanical movers that allowed the detector
element  to  be  moved  remotely  through  the  scattered  beam.   In  one  application  this
enabled us to determine the full spatial profile of the scattered beam intensity.  Of course,
the same result could also have been achieved by using an array of detectors and 
light--guides.   However, this would have required periodic cross--calibrations of the responses
of  the  various  detectors.  The  use  of  a  single,  or,  at  most,  a  small  number  of 
scanning
detectors not only avoided this chore, but also facilitated the removal of the detectors
from the scattered beam profile during data acquisition with other detectors.

 These detectors were developed for SLAC experiment E158 \cite{e158} whose objective was to
determine the weak mixing angle by the measurement of parity--violation in the M{\o}ller
scattering  of  polarized  electrons  from  electrons  in  a  liquid  hydrogen  target.   This
experiment  employed  a  forward  spectrometer,  and  the  polarization  asymmetry  of  the
M{\o}ller cross section was determined by flux--integration of the forward--scattered electron
beams  of  different  helicities.   Due  to  the  design  of  the  E158  spectrometer,  scattered
electrons were constrained to two forward--directed circular bands between 4.4 and 7.5 mr
surrounding  an  intense  background  of  forward  photons  generated  by  bremsstrahlung
within the target.  Near the main E158 calorimeter detector, the radial distribution of the
scattered electron ranged from 18--35 cm, with M{\o}ller electrons confined to an inner band
of  radius  18--24  cm  containing  about  $2*10^7$  electrons  per  beam  pulse.   Outside  the
M{\o}ller band was a concentric band of electrons produced by the scattering of the incident
electrons off the protons in the hydrogen target.
\section{Small--probe Cherenkov detectors}
\label{sec2}
\subsection{General Discussion}
\label{sec2.1}
The  primary  function  of  these  detectors  was  to  carefully  map  the  relative  spatial
distribution of the scattered beam before it was absorbed in the M{\o}ller calorimeter.  This
mapping constituted the principal means of ensuring that the beam was centrally aligned
at the detector, and that the inner ring of M{\o}ller electrons was clearly separated from the
outer ring of e--p electrons. These measurements were critical  for tuning the spectrometer during
detector commissioning.

Figure 1 illustrates the essential features of these detectors.  The active element was a
$5 *5*20$ mm$^3$ piece of fused silica.  This was oriented at an angle of 45$^\circ$ with respect
to the incident beam in order to match the Cherenkov angle for relativisitic electrons in
fused silica. In order to increase the sensitivity of the device, a rhombic $5*15*15$ mm$^3$
tungsten  pre--radiator  was  mounted  on  the  upstream  side  of  the  radiator.   Cherenkov
photons emerging from the end of the radiator were reflected along a tube lined with a
reflective Alzak \cite{Alzak} foil until they were detected by a photomultiplier tube (PMT).
For each beam pulse the total charge collected on the PMT anode was digitized by the 
LeCroy 2249W ADC.

\begin{figure}[h] 
\epsfig{file=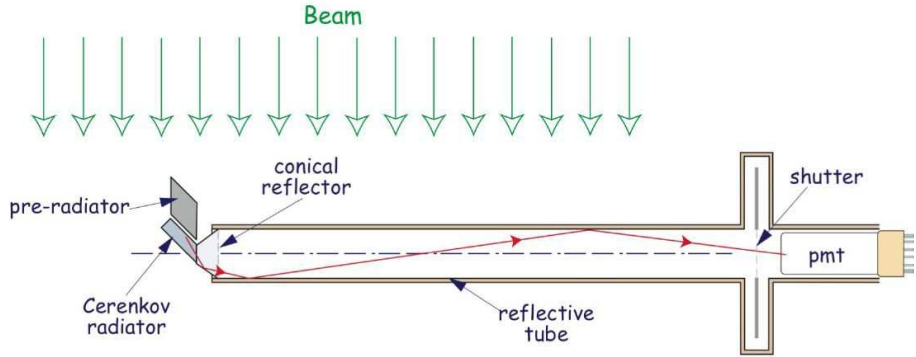,height=3.0in,angle=0,bbllx=-50,bblly=0,bburx=800,bbury=500,clip=,silent=}
\caption{Schematic  layout  for  measuring  the  spatial  distribution  of  a  broadly  
distributed  
intense  beam. Cherenkov photons produced in the radiator are reflected to a remotely located 
photomultiplier tube (PMT) by means  of  a  air--core  light  guide. By  closing  the  shutter,  
a measurement  can  be made  of  the  background produced by ionizing radiation striking the PMT.}
\label{fig1}
\end{figure}
\subsection{Design Considerations}
\label{sec2.2}
Extensive Monte--Carlo simulations were made to optimize the design of the detector.
These  revealed  that  even  though  the  fused  silica  radiator  was  aligned  to  match  the
opening  angle  of  the  Cherenkov  light  cone,  the  overwhelming  majority  of  detected
Cherenkov photons were multiply reflected inside the radiator.   Because not all of these
reflections proceeded by total internal reflection, wrapping the radiator in reflective foil
increased the sensitivity of the device by about 10\%.

 Due to the multiple reflections inside the radiator, Cherenkov photons emerge from the
end of the radiator with a large range of angles. Simulations showed that a reflector cone
surrounding the end of the radiator would be effective in deflecting Cherenkov photons
towards the PMT, thereby increasing the sensitivity of the device.

Two provisions were made to help identify and determine experimental backgrounds.
First, as indicated in Fig. \ref{fig1}, an electromechanical shutter was installed immediately in
front of the PMT. With the shutter open, reflected Cherenkov photons could enter the PMT
and  interact  in  its  photocathode.   When  shut,  however,  the  shutter  intercepted  those
photons permitting a measurement of the background due to ionizing radiation hitting the
PMT.   The second provision, not indicated in Fig. \ref{fig1}, was the capability of being able to
move the pre--radiator from the upstream to the downstream side of the radiator.   This
allowed us to ensure that the signal was due to the beam directly striking the pre--radiator,
rather than from secondary interactions of the beam in other nearby components of our
experimental setup.

\begin{figure}[h] 
\epsfig{file=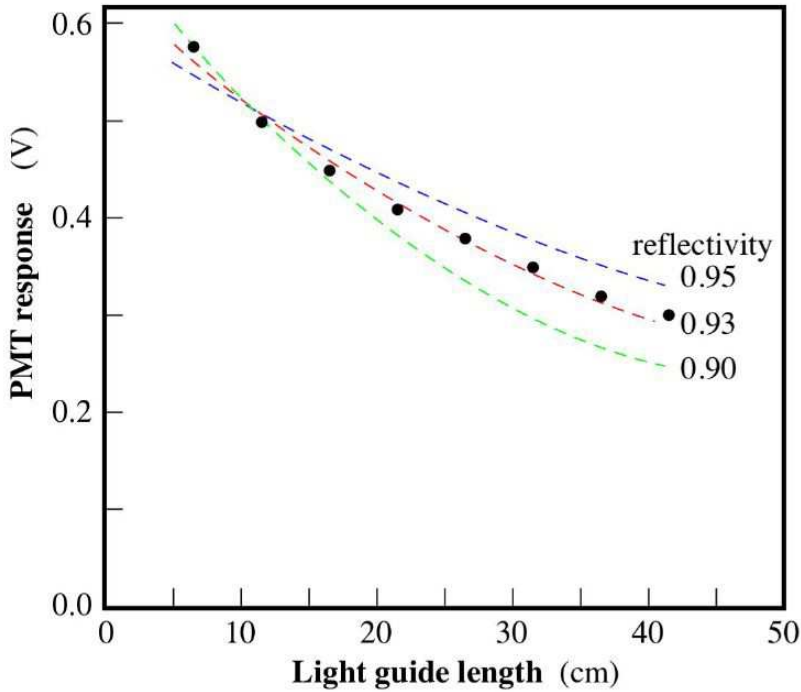,height=4in,angle=0,bbllx=-40,bblly=0,bburx=800,bbury=500,clip=,silent=}
\caption{ Relative transmission of light through a 19 mm diameter tube lined with reflective Alzak 
material. The light from a blue--light emitting diode into a 30$^\circ$ cone was reflected from 
various lengths of the Alzak tube and measured using a photomultipler tube (PMT).  The points are 
data; the lines represent the results of simulations with different reflectivity values.}
\label{fig2}
\end{figure}

The key variables affecting the sensitivity were the internal diameter and length of the
reflector  lined  tube,  and  the  properties  of  the  radiator,  pre--radiator,  and  reflecting
surfaces.   One concern was that rolling the Alzak sheet to form a small--radius tubular
light--guide  could  degrade  the  95\%  reflectivity  specified  for  our  Alzak  material.
However,  as  indicated  in Fig. \ref{fig2},  which  compares  measured  and  simulated \cite{geant}
photon transmission in a 19 mm diameter tube of varying length, this concern was unfounded: a
reflectivity of 93\% gives excellent agreement between experiment and theory.

\begin{figure}[h] 
\epsfig{file=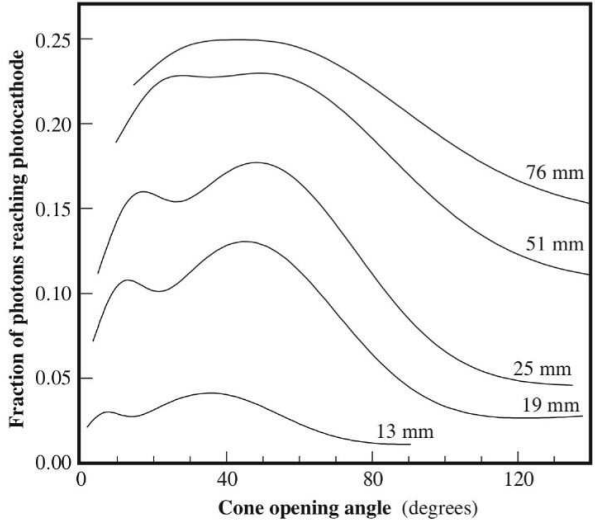,height=3.0in,angle=0,bbllx=-150,bblly=0,bburx=600,bbury=500,clip=,silent=}
\caption{ Dependence of simulated light transmission on opening angle of mirror cone and 
diameter of a 50 cm long reflective tube.  The calculations are for the wavelength range 
350 -- 440 nm.}
\label{fig3}
\end{figure}

 Figure \ref{fig3} shows  the  simulated Cherenkov light collection as a function of light guide
diameter and cone opening angle.   Only photons in the wavelength range 350 -- 440 nm
were considered, corresponding to a quantum efficiency of  25\% for most photocathode
materials.  The  photons  were  assumed  to  undergo specular  reflection  within  the  
light--guide cone and tube until they reached the PMT entrance window located 50 cm from the
radiator.   Photons that struck the PMT window at large angles of incidence failed to be
transmitted through to the photocathode.

   As expected, for small tube diameters the calculated collection efficiency drops due to
the large number of reflections required for propagation along the full length of the tube.
The light collection generally has a broad maximum for cone opening angles of 
35$^\circ$--60$^\circ$.

\begin{figure}[h] 
\epsfig{file=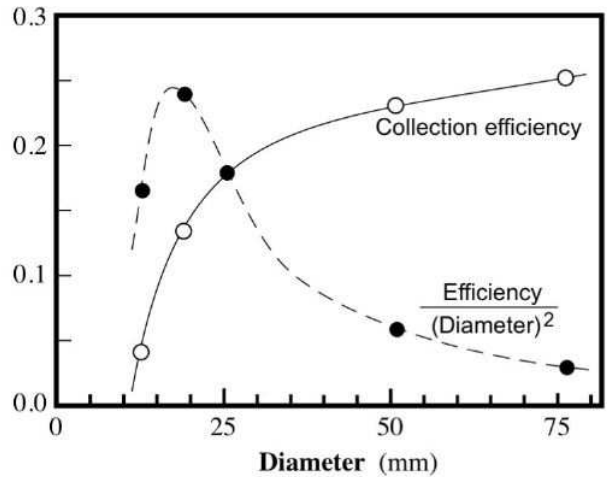,height=3.5in,angle=0,bbllx=-60,bblly=0,bburx=800,bbury=400,clip=,silent=}
\caption{Open points: Collection efficiency versus tube diameter for light propagation along a 
50 cm--long tube. Solid points: Collection efficiency divided by tube diameter squared, 
arbitrarily normalized. This ratio constitutes  a  figure--of--merit  for  the  
signal--to--background  
ratio,  where  the  background  is  produced  either within the light guide tube, or in the 
PMT entrance window.}
\label{fig4}
\end{figure}

Although  large  matching  light  guide  and  PMT  diameters  favor  high  collection
efficiency, they also make the detector more susceptible to background arising from the
flux of particles traversing the light guide between the radiator and PMT photocathode, as
well as to secondary radiation interacting directly in the PMT photocathode. In part, the
corresponding signal--to--background ratio may be quantified by the ratio of the collection
efficiency to the volume of the light guide or PMT window. An equivalent figure--of--merit,
indicated in Fig. \ref{fig4}, is the collection efficiency divided by the square of the 
light guide or
PMT diameter. The corresponding ratio is seen to be optimal for a PMT diameter close to
that of the 19 mm diameter tube used in our design. Somewhat different models for
the background generation do not significantly alter this conclusion.

   Cherenkov and scintillation photons produced in the air contained within the light guide
constitute most of the background.  In contrast to the Cherenkov photons that are radiated
at an angle of only 1.4$^\circ$ in air relative to the initial electron direction, 
scintillation photons
are isotropically distributed. Using results of ref. \cite{bert}, one can estimate that a minimum
ionizing particle passing through one cm of air produces about 0.0066 photons/sr in the
wavelength range matching the photomultiplier cathode sensitivity.   Assuming a 2 sr
solid angle of photon collection in the 19 mm diameter reflective--tube and the 1.5 cm
average path of the electron traversing it, yields one scintillation photon reaching the PMT
per  about  20  passing  electrons.  This  number  should  be  compared  with  about  200
Cherenkov photons reaching the PMT per electron passing through the radiator, as will be discussed,
and is shown in Fig. \ref{fig6}.
Therefore under condition of uniform illumination, each cm of the tube would contribute
about 0.06\% to the signal from the Cherenkov light produced within the radiator.   This
number is sufficiently low to assure that even with an electron flux density varying by
two orders of magnitude, the scintillation contribution from air in the light guide will not
exceed a few percent.

   Due  to  the  small  angle  of  Cherenkov  photons  produced  in  air,  Cherenkov  radiation
produced within the light guide was of little concern over most of the length of the light
guide tube: these photons just reflect back and forth, transverse to the tube axis.

\begin{figure}[h] 
\epsfig{file=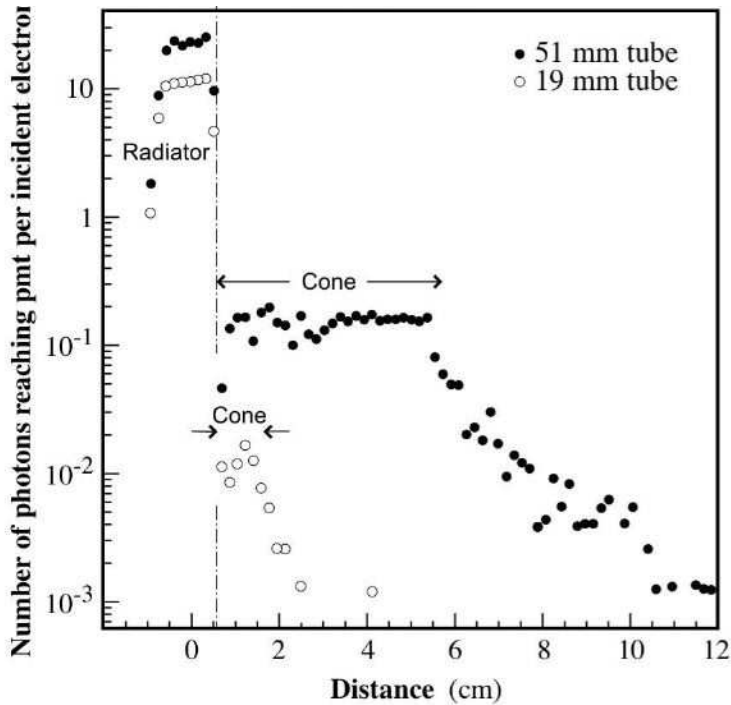,height=4.0in,angle=0,bbllx=-40,bblly=0,bburx=800,bbury=430,clip=,silent=}
\caption{Results of simulations showing the position of origin of Cherenkov photons that reach a 
PMT at the end  of  a  50  cm--long  light  guide  tube  for  two  different  tube  diameters.   
The  reflectivity  of  the  tube  is assumed to be 93\%.  The solid (open) points are for a tube 
of internal diameter of 51 mm (19 mm).} 
\label{fig5}
\end{figure}

\begin{figure}[h] 
\epsfig{file=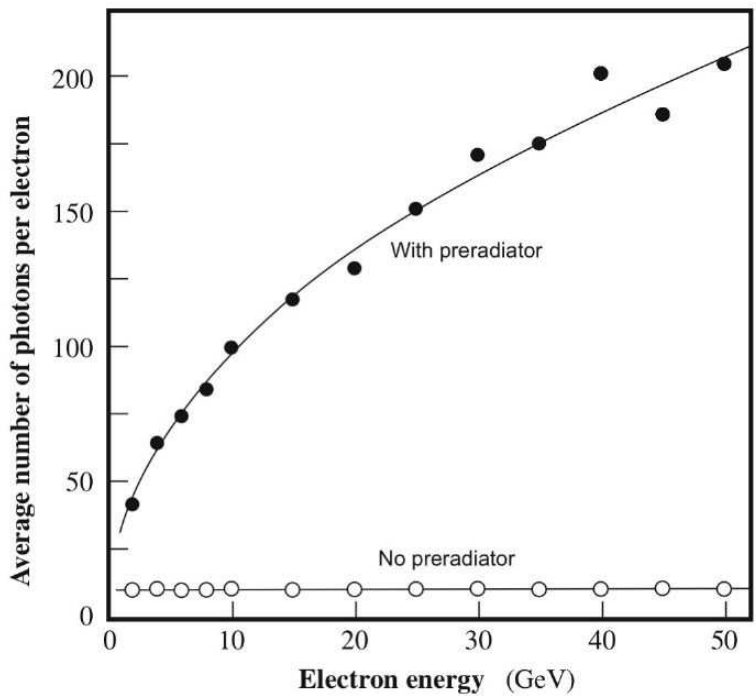,height=4in,angle=0,bbllx=-50,bblly=0,bburx=800,bbury=500,clip=,silent=}
\caption{Results of simulations for a 50 cm--long, 19 mm--diameter tube, with and without a five 
radiation--length pre--radiator.  The ordinate is the number of photons reaching the PMT 
per incident electron striking the pre--radiator. The solid line shows the second order 
polynomial fit to the data.}
\label{fig6}
\end{figure}

On the other hand, Cherenkov photons produced within the reflective cone adjacent
to  the  radiator  are  not  similarly  trapped;  such  photons  may -- with  just  one  or  two
reflections -- enter  the  PMT  window.   Figure \ref{fig5}  shows  the  results  of  
simulations  of
Cherenkov  production  in  the  radiator  and  in  the  air--core  light  guide  under  uniform
electron  flux  irradiation  for  two  light--guide  diameters,  19  and  51  mm,  assuming
identical $5*5*20$ mm$^3$ radiators of fused silica.  As indicated, the larger diameter tube
has a collection--efficiency for light produced in the radiator that is better by a factor of
two.   However, due to the much smaller air volume within its cone, the background
produced within the light--guide of the 19 mm diameter tube is smaller by about two
orders--of--magnitude, a striking improvement of scanner flux profile resolution, well
worth the cost of the reduced signal in the narrower tube.

\begin{figure}[h] 
\epsfig{file=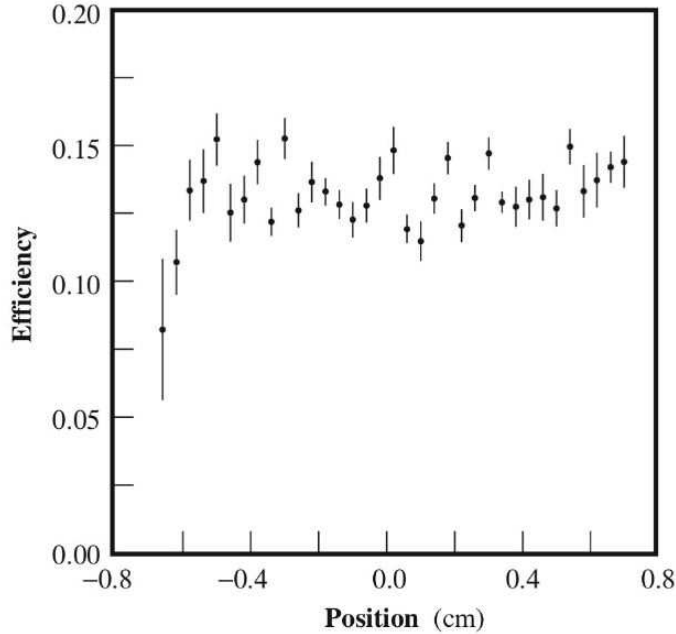,height=4in,angle=0,bbllx=-90,bblly=0,bburx=800,bbury=500,clip=,silent=}
\caption{Simulated light  collection  efficiency  as  a  function  of  interaction  position  in  
quartz  radiator.  The  position shown is the projection along the light--guide axis.}
\label{fig7}
\end{figure}

 As indicated in Fig. \ref{fig6}, the addition of the tungsten pre--radiator immediately upstream
of the fused quartz radiator makes the detector much more responsive.   The signal--to--background  
ratio  is  similarly  improved.   In  addition,  the  pre--radiator  enhances  the
directionality of the detector response, making it relatively insensitive to electron 
back--splash and other off--axis backgrounds. Finally, results of simulations shown in  
Fig. \ref{fig7} demonstrate the uniformity  
of the light collection as a function of the Cherenkov production position within the radiator.

\subsection{Construction}
\label{sec2.3}
\begin{figure}[h] 
\epsfig{file=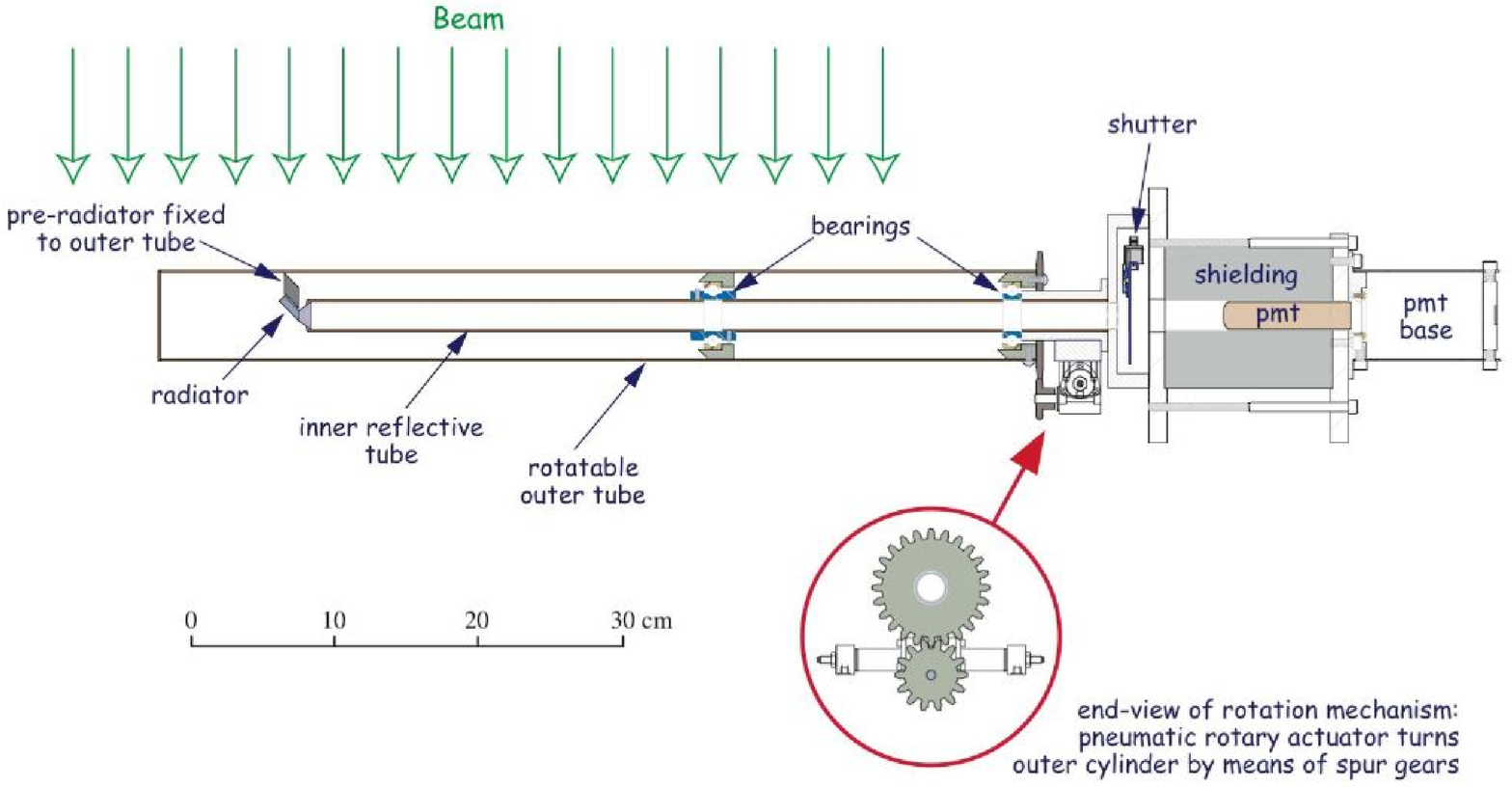,height=3.4in,angle=0,bbllx=0,bblly=0,bburx=800,bbury=500,clip=,silent=}
\caption{Final design for measuring spatial distribution of an intense, broadly distributed 
electron--beam. In order to help identify and quantify background, a rotary actuator can move 
the pre--radiator to a position immediately downstream of the fused quartz radiator. The 
background 
is also reduced by surrounding the PMT with sintered tungsten shielding.}
\label{fig8}
\end{figure}

Figure \ref{fig8} shows the final version of the scanning Cherenkov detector that was built for
experiment E158.    Cherenkov light was produced in a fused silica radiator measuring 
$5*5*20$ mm$^3$, 
canted at an angle of 45$^\circ$ to the incoming electron beam.  As noted above,
the 45$^\circ$ alignment of the quartz with respect to the beam served to optimize the response
of  the  Cherenkov  detector  to  the  incoming  beam  and  reduce  sensitivity  to  background.
Due to the high intensity of the scattered beam, the PMT that collects this Cherenkov light
was situated just outside the E158 phase--space, about 50 cm from the radiators.

 The inner tube was aluminum, with an inner diameter of 20.5 mm and a wall thickness
of 0.08 mm. It was lined with a rolled sheet of Alzak--surfaced aluminum. The sheet we
used was difficult to roll because it was 0.3 mm thick; if available, thinner stock would be
preferred.

The electromechanical shutter \cite{uni} had a circular aperture of 25.4 mm. Normally the
shutter was open, but when closed it allowed us to check for background generated by
stray ionizing radiation directly hitting the 19 mm diameter PMT \cite{emi}. In order to reduce
this background the sides of the PMT were shielded by 4 cm of tungsten.
\subsection{Motion}
\label{sec2.4}
\begin{figure}[h] 
\epsfig{file=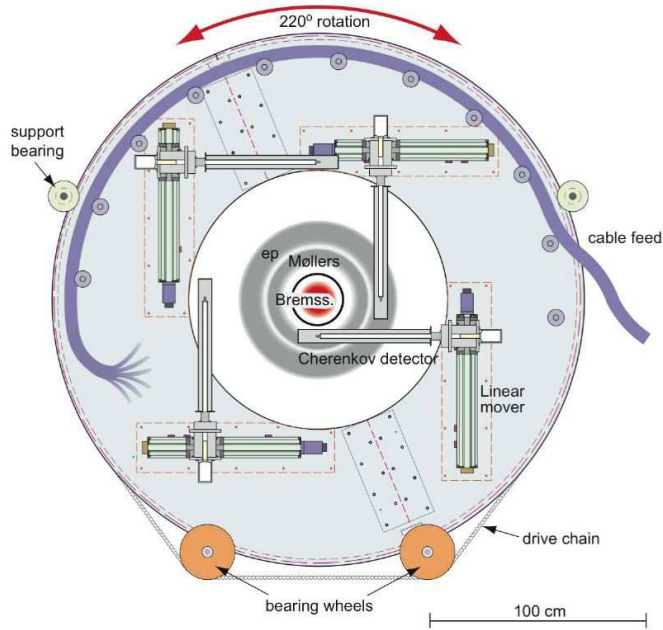,height=3.6in,angle=0,bbllx=-100,bblly=0,bburx=900,bbury=600,clip=,silent=}
\caption{Arrangement of four Cherenkov detectors on rotatable annulus. Linear movers translate the 
detectors radially, while the annulus can be rotated 220$^\circ$ by means of chain embedded in 
its rim. The annulus was supported  on  two  large  brass  bearing  wheels,  and  stabilized  in  
the  vertical  plane  by  smaller  bearings  at about 2/3 of the annulus height. The entire system 
was enclosed by a frame attached to a moveable cart that also provided earthquake protection.  
High--voltage, signal, and motion control cables were bundled on both sides around the perimeter of 
the annulus, and held on hangers at 15$^\circ$ intervals so as to accommodate the cable transport 
through the full 220$^\circ$ rotation of the annulus.  During experimental runs a tungsten--lead
collimator was installed to intercept electrons scattered from the protons.}
\label{fig9}
\end{figure}

As described above, in experiment E158 the electrons scattered from the protons and
from  the  electrons  in  the  hydrogen  target  formed  two  concentric  circular  bands
surrounding an intense forward--directed bremsstrahlung background.   In order to scan
the two bands, Cherenkov detectors were positioned 90$^\circ$ apart on a rotatable annulus of
outer diameter 246 cm as shown in Fig. \ref{fig9}.  The detectors were mounted on four linear
movers \cite{parker} so that they could be moved radially relative to the beam axis.  The annulus
could be rotated through an angle of 220$^\circ$ thus allowing the entire profile of the scattered
electrons to be measured.  In order to avoid beam interception during the experimental
runs,  the  Cherenkov  detectors  were  moved  to  a  radial  position  beyond  the  scattered
electron beam.

The annulus \cite{annulus} had an outside aluminum rim of 44 mm radial thickness and 108 mm
depth onto which was welded a 6 mm thick circular aluminum plate. For ease of transport
and flexibility in assembly, it separated into two halves joined together by fish plates and
by tension blocks welded under the rim.  The annulus was entrapped and supported by a
pair of large brass bearing wheels, and was rotated by means of turning a sprocket wheel
that engaged a roller chain fixed in a groove machined into the edge of the rim.   The
sprocket  wheel  itself  was  driven  through  a  20:1  reduction  gear  by  a  stepping  motor
mounted on the support frame for the annulus.   When assembled at SLAC with its full
complement of detectors, the roundness of the annulus was measured to be better than
0.25 mm.

The annulus rotation and the linear translation of the Cherenkov scanners were remotely
controlled by means of a LabView Flexmotion \cite{lv} application. The Cherenkov scanners
were equipped both with optical rotary encoders for the motion control, and with linear
potentiometers \cite{pot} for providing position information into the data acquisition system.
\subsection{Performance}
\label{sec2.5}
\begin{figure}[h] 
\epsfig{file=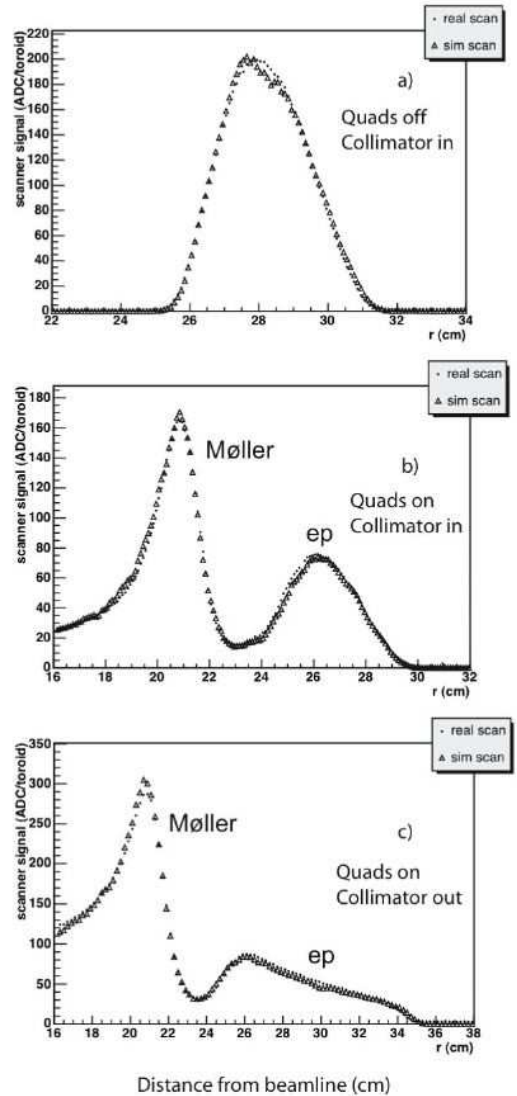,height=6in,angle=0,bbllx=-100,bblly=0,bburx=800,bbury=650,clip=,silent=}
\caption{Radial  response  of  scanning  Cherenkov  detector,  and  comparison  to  
normalized  Monte  Carlo prediction for the 150 cm long liquid hydrogen target. Panels a) and b)
show measurements (dots) and simulations (triangles) with quadrupole magnets switched off and on, 
respectively, and a collimator inserted between the dipole and the quadrupole magnets. This 
collimator selected electrons scattered at a fixed azimuthal angle. The focusing effect of the 
quadrupole magnets leading to a separation of electrons scattered from electrons and protons is 
clearly seen. Panel c) shows scattered beam profile scan during a typical production run.}
\label{fig10}
\end{figure}

 The scanner system worked well throughout the two--year period of this experiment.
Apart from the commissioning periods when it was used extensively, the scanner was
typically  used  for  beam  profile  checks  about  twice  each  day.  Despite  the  large
integrated luminosity of this experiment, no degradation was observed in the scanner
performance.  Fig. \ref{fig10} shows results of radial scans during the E158 magnetic 
spectrometer commissioning, and a typical scan during the production run. They are compared with 
the results of the GEANT Monte Carlo simulations of the spectrometer response. Very good 
agreement between measured scans and simulations gives account of a good control in simulations 
of all effects influencing the spectrometer performance.  
\section{Cherenkov Polarimeter}
\label{sec3}
\subsection{Introductory Remarks}
\label{sec3.1}
In experiment E158 the polarization of the incident electron beam was determined by
measuring M{\o}ller scattering from magnetized iron foils mounted at 20$^\circ$ with respect to
the beam and ranging in thickness between 20 and 100 $\mu$m. For such thin targets, the
M{\o}ller count rate was reduced by a factor of  200 compared to the rate from the 1.5 m
long  liquid  hydrogen  target  used  in  the  experiment.  In  order  to  eliminate  the e--p
background and define a certain kinematics acceptance, a collimator was used which
limited the scattered beam to an area of 7 x 1 cm$^2$ in the detector plane. The M{\o}ller
calorimeter  was  not  suitable  for  polarimetry  measurements  because  its  size  and
geometry did not match the collimator geometry. On the other hand, the reduction of the
scattered flux intensity would have made it difficult to acquire the desired precision with
the small scanning Cerenkov detector described in section \ref{sec2}.

\begin{figure}[h] 
\epsfig{file=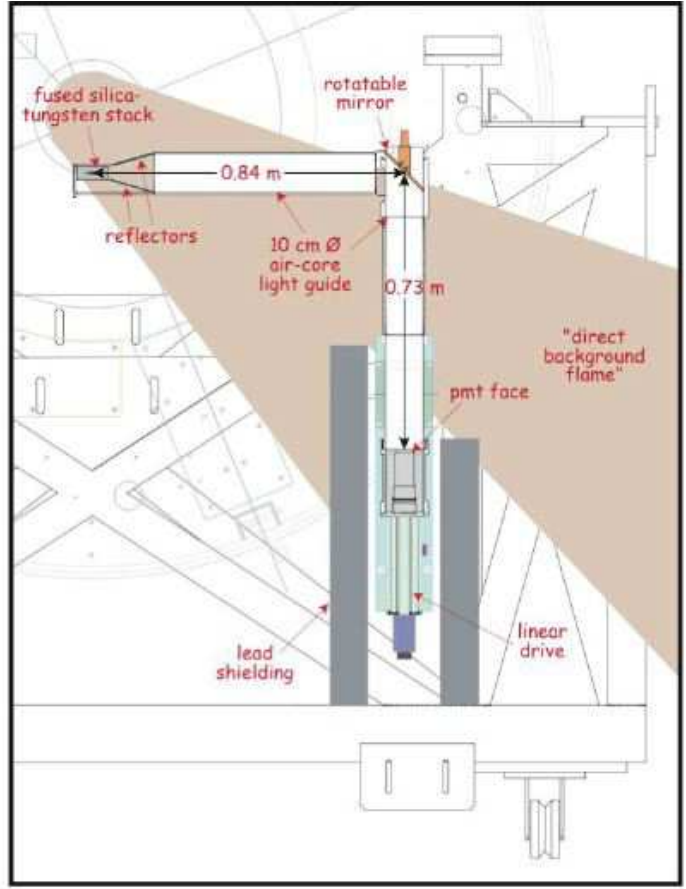,height=5.0in,angle=0,bbllx=-50,bblly=0,bburx=800,bbury=400,clip=,silent=}
\caption{Air--core Cherenkov detector installed on detector cart and used for polarimetry in 
experiment E158 at  SLAC.   Driven  by  a  linear  mover,  the  fused  silica--tungsten  detector  
head  can  be  moved  vertically through a band of scattered electrons beneath the beam--axis. 
Ten cm of lead shielded the PMT from line--of--sight background produced in the vicinity of the 
beam--pipe.}
\label{fig11}
\end{figure}

 Hence a second air--core detector was designed that featured a much larger and thicker
active element.  Nevertheless, the overall concept for this polarimeter was similar to that
of the smaller air--core Cerenkov scanners.  As shown in Fig. \ref{fig11}, the air--core light guide
of the polarimeter was a 10 cm internal diameter tube whose internal surface was lined
with Alzak \cite{Alzak}. Cerenkov photons were reflected around a 90$^\circ$ angle in the tube 
by means
of an Alzak--surfaced rotatable mirror that could be set to bisect the angle between the
horizontal and vertical tubes.  This arrangement has two benefits.  First, by rotating the
mirror 180$^\circ$ about a vertical axis, Cherenkov photons that would normally be reflected
towards the PMT were obstructed.  This allowed a measurement of the background due to
radiation interacting directly in the pmt.  Effectively then, the rotatable mirror served the
same function as the shutter in the scanning Cherenkov detector. The second feature, a
significant improvement over the more compact Cherenkov scanner, is that the 90$^\circ$ angle
in the light--guide permitted the photocathode of the PMT to be shielded from line--of--sight
background originating in the vicinity of the beam pipe. In our case the shielding was
provided by 10 cm of lead surrounding the detector.

 The active element of the polarimeter consisted of alternating layers of fused silica
and tungsten of total thickness of 13 radiation lengths. 
\subsection{Design}
\label{sec3.2}
\begin{figure}[h] 
\epsfig{file=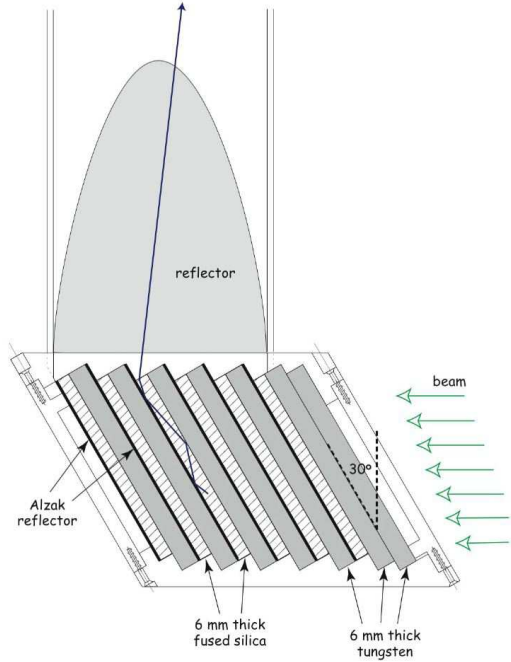,height=3.6in,angle=0,bbllx=-250,bblly=0,bburx=800,bbury=700,clip=,silent=}
\caption{Arrangement of tungsten and fused silica layers in the active element of the air--core 
polarimeter.
Cherenkov photons generated by the electromagnetic shower reflect along the fused silica layers 
and emerge
from the top.   In the plane normal to this view, the photons are reflected close to the axis of 
the air--core
light guide by two planar semi--elliptical reflectors, one above and the other below the detector 
elements.
Alzak foils placed against each fused silica plate increased the response of the detector by 
about 30\%.}
\label{fig12}
\end{figure}

\begin{figure}[h] 
\epsfig{file=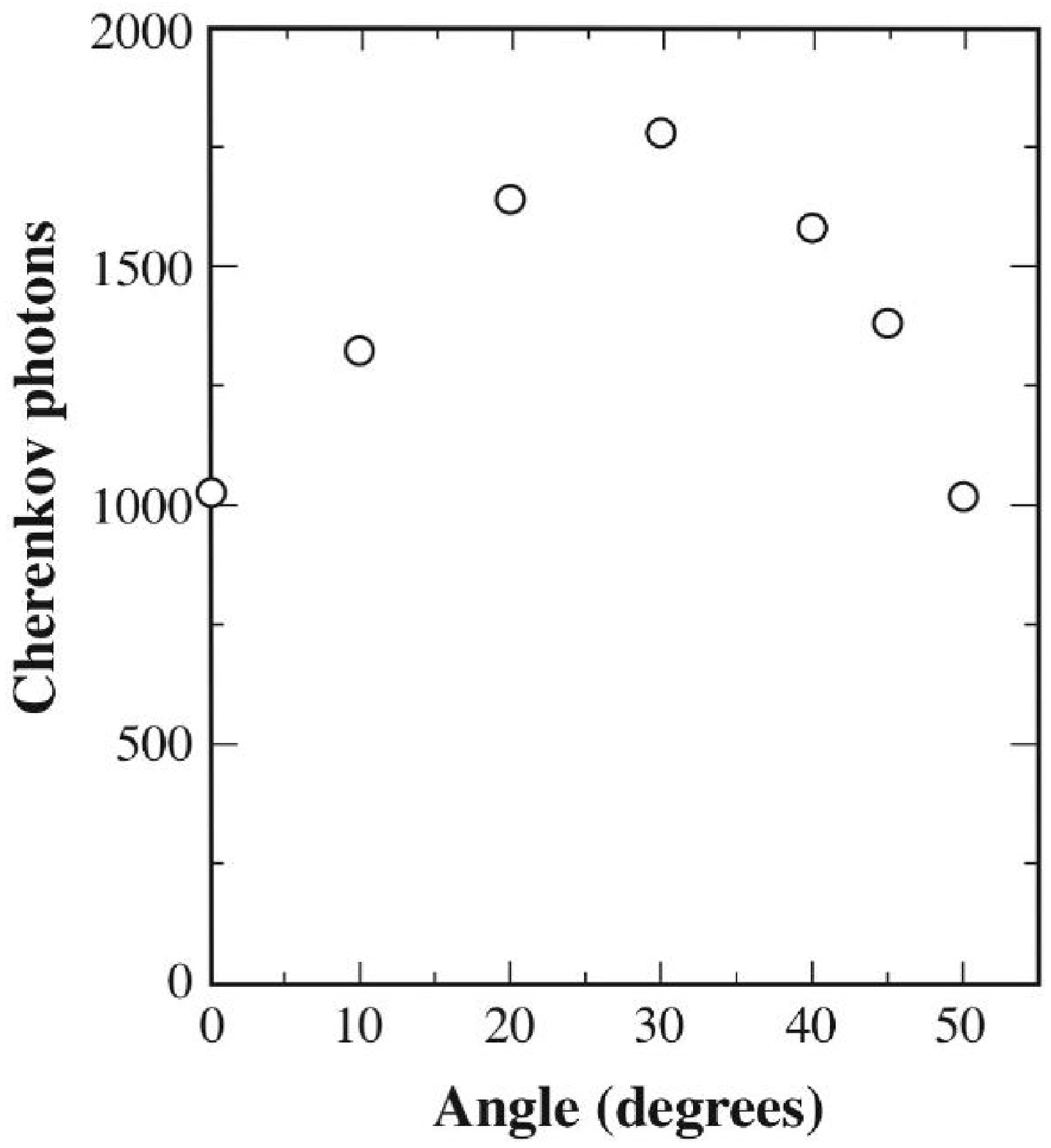,height=3.8in,angle=0,bbllx=-130,bblly=0,bburx=800,bbury=550,clip=,silent=}
\caption{Calculated number of Cherenkov photons in the air--core light--guide, for a single 22 GeV 
electron
showering in the detector stack.  The response is calculated as a function of the angle between 
the incident beam and the normal to the fused silica--tungsten stack.}
\label{fig13}
\end{figure}

The transverse size of the active element was optimized to match the 7 x 1 cm$^2$ image
produced by special spectrometer optics used in the polarimetry measurements. In order
to  allow  for  the  transverse  development  of  the  electromagnetic  shower  within  the
detector, the size of the fused silica--tungsten stack was made slightly larger than this.
Each of the six fused silica plates measured 9.3 x 3.5 x 0.6 cm$^3$ , and the tungsten plates,
7 in all, 9.8 x 3.5 x 0.6 cm$^3$ . As shown in Fig. \ref{fig12}, this stack was tilted relative 
to the
beam  by  30$^\circ$,  an  angle  calculated  to  optimize  the  response  of  the  detector.   The
calculated dependence on angle is shown in Fig. \ref{fig13}.
Extensive GEANT \cite{geant} Monte--Carlo calculations were made to optimize the details of
the detector.  These calculations guided the choice of the number and thicknesses of the
fused  silica  and  tungsten  layers,  as  well  as  their  orientation  relative  to  the  
beam. In
addition to the Alzak lining of the  light--guide tube, the simulations also favored the use
of reflective surfaces elsewhere in the detector.  For example, calculations showed that
reflective foils laid against each fused silica plate increased the response of the detector
by about 30\%.  The simulations also showed the efficacy of two planar reflectors, 
semi--elliptical in shape, installed above and below the detector element at an angle 
of 15$^\circ$ to
the tube axis.  Cherenkov photons emerging from the fused silica layers were reflected by
these Alzak reflectors such that they followed paths more parallel to the axis of the 
air--core  light  guide.   This  reduced  the  number  of  reflections  each  ray  underwent,  thus
improving the light collection.

\begin{figure}[h] 
\epsfig{file=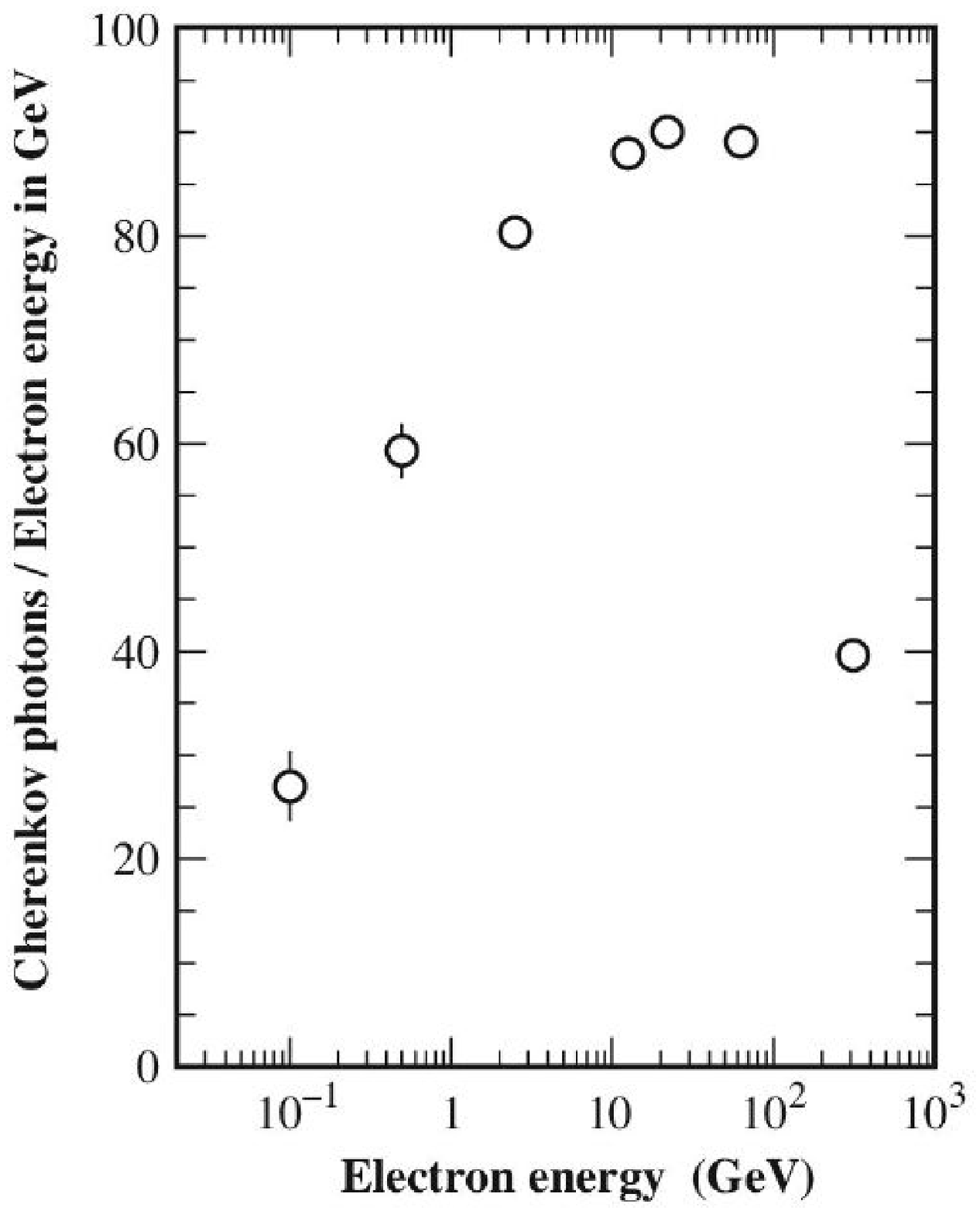,height=3.8in,angle=0,bbllx=-130,bblly=0,bburx=800,bbury=550,clip=,silent=}
\caption{ Monte Carlo results for dependence of detector response on electron energy.  The response is quite
linear for the 20--30 GeV energy--range of M{\o}ller--scattered electrons in experiment E158.}
\label{fig14}
\end{figure}

\begin{figure}[h] 
\epsfig{file=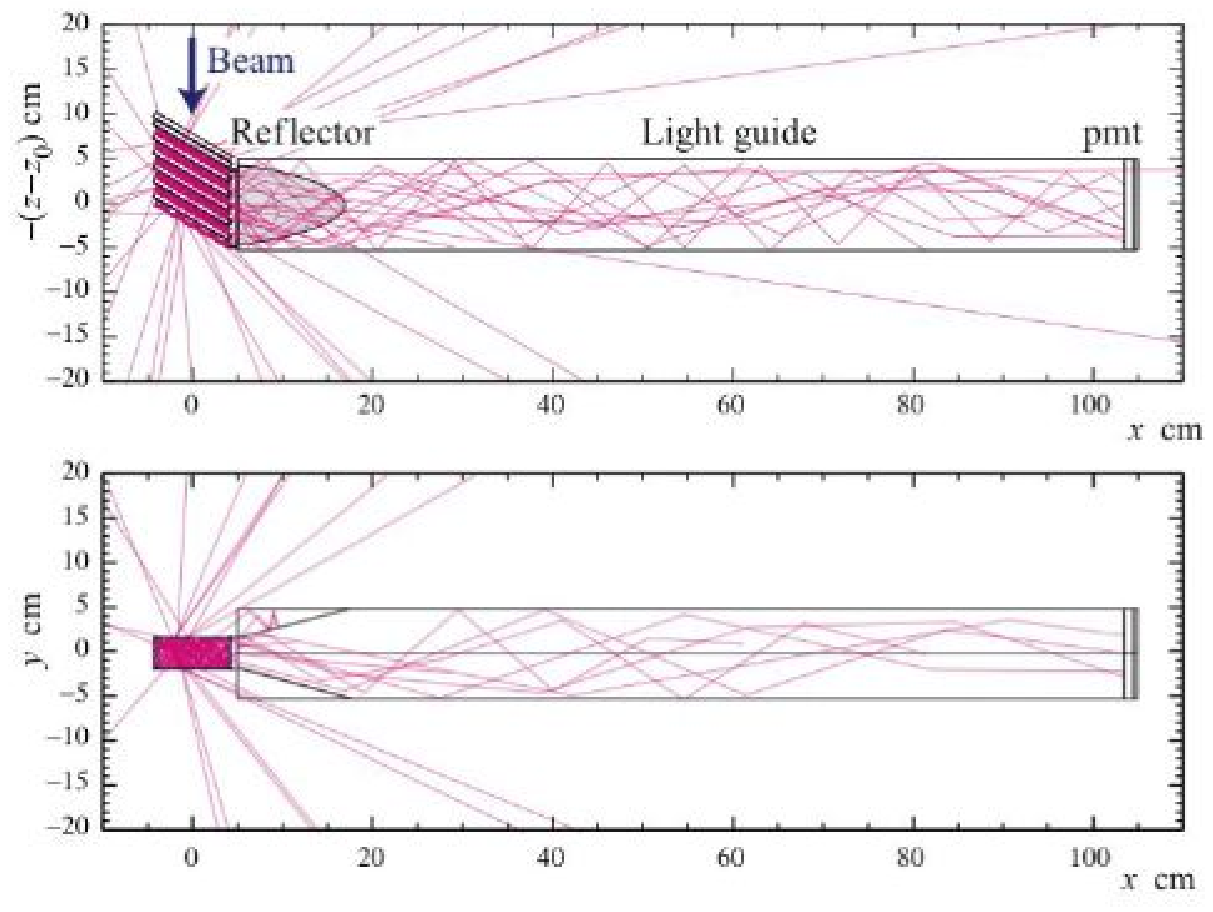,height=4.5in,angle=0,bbllx=10,bblly=0,bburx=800,bbury=280,clip=,silent=}
\caption{Monte Carlo simulation showing 0.5\% of Cherenkov photons produced by the showering of a
single 22 GeV electron in the air--core polarimeter.}
\label{fig15}
\end{figure}

 The adopted design provided not only large signal strength, but also the necessary
energy resolution to discriminate against low energy backgrounds.  As indicated in 
Fig. \ref{fig14}, the response of the calorimeter is also reasonably linear in a range of 
10--100 GeV,
though the relative signal drops considerably for low and for high energies. The drop at
low energies aids in reducing background.
Figure \ref{fig15} shows two perspectives of the simulated response of the detector to the
absorption of a 22 GeV electron. No shower products are shown other than the Cherenkov
photons.  Moreover, for clarity, only 0.5\% of the expected number of photons were
simulated.
The simulations, for a 1 m long straight air--core light guide with a 10 cm internal
diameter and a 10 cm phototube indicated that about 1800 Cherenkov photons reach the
PMT for every 22 GeV M{\o}ller electron absorbed in the detector. Because the constructed
light guide was 1.6 m long with a 90$^\circ$ bend, and a 5 cm Hamamatsu PMT \cite{PMT} was used,
in practice, the yield was considerably lower.   Moreover, the window material of the
PMT was {\em not} of a type that favors high ultraviolet transmission, and hence a fraction of
the Cherenkov photons were absorbed before reaching the PMT photocathode.  However,
these  losses  were  considered  acceptable  in  view  of  the large  flux  of  
M{\o}ller--scattered  electrons incident upon the polarimeter (a few hundred electrons per 
10$^{11}$ electrons in the beam pulse).
\subsection{Some practical information}
\label{sec3.3}
 The rotatable  mirror  was  fabricated  by  bolting  a  6.3  mm  thick  brass  plate  to  the
45$^\circ$--beveled end of a 33 mm diameter aluminum shaft.  By turning the shaft in a lathe,
the plate was machined to the required elliptical profile.  An Alzak foil was fixed to the
brass  plate  to  provide  the  reflective  surface.  The  assembly  was  rotated  by  a  180$^\circ$
pneumatic actuator \cite{bimba}, and limit switches were added to indicate the mirror position.

When not in use, the polarimeter was moved outside of the scattered beam region.  As
for the Cherenkov scanners, this motion  through a radial distance of 0.5 m, was provided
by a linear mover \cite{parker} equipped with limit and home switches, and
rotary  and  linear  position  encoders.   The  square  rails  of  the  linear  mover  were  
well--suited for resisting the appreciable torque imposed by the 5 kg of tungsten supported a
distance of 85 cm off to the side of the mover's axis.

As  with  the Cherenkov scanners, the motion of the polarimeter was controlled by a
National Instruments Flexmotion  \cite{lv}  with its linear potentiometer providing position
information to the data acquisition system.
\subsection{Performance}
\label{sec3.4}
\begin{figure}[h] 
\epsfig{file=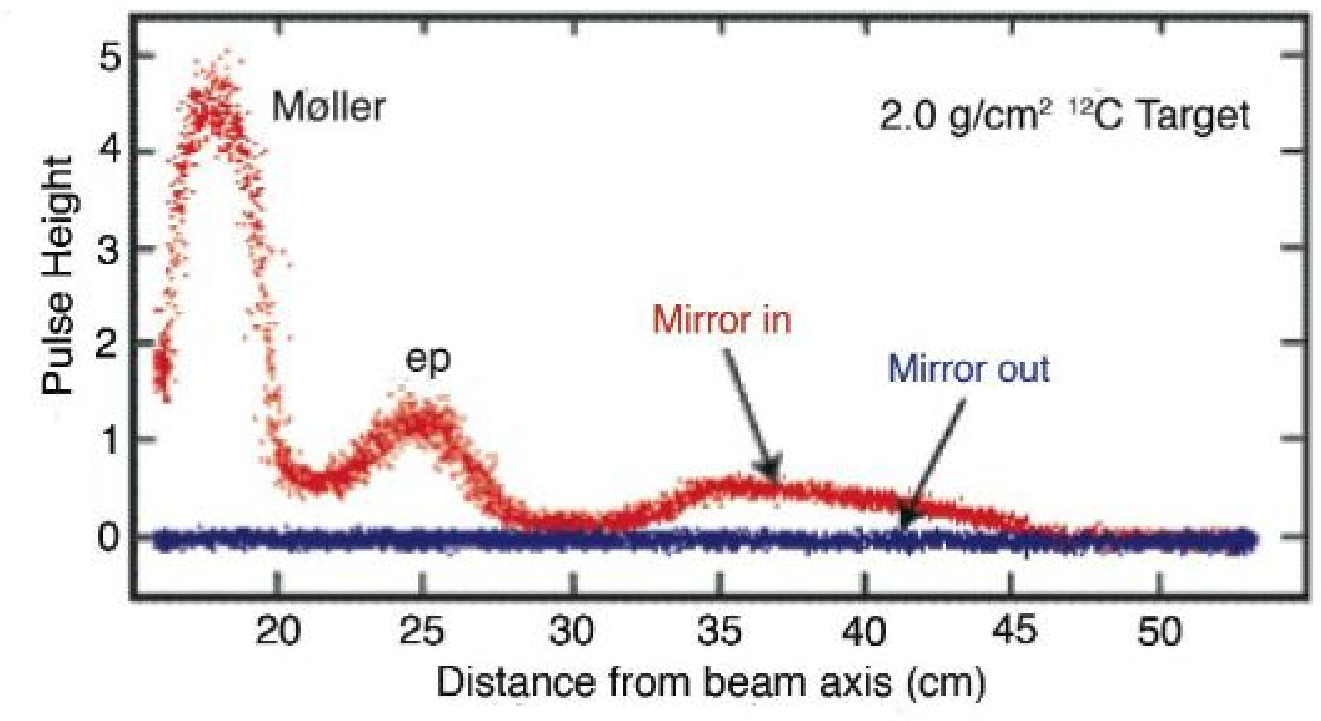,height=3.5in,angle=0,bbllx=10,bblly=0,bburx=400,bbury=250,clip=,silent=}
\caption{Polarimeter response measured as a function of radial distance from the beam axis. In 
the ``Mirror in'' position the mirror reflects Cherenkov light produced in the calorimeter to the 
PMT. In the ``Mirror out'' position this light is obstructed.}
\label{fig16}
\end{figure}

\begin{figure}[h] 
\epsfig{file=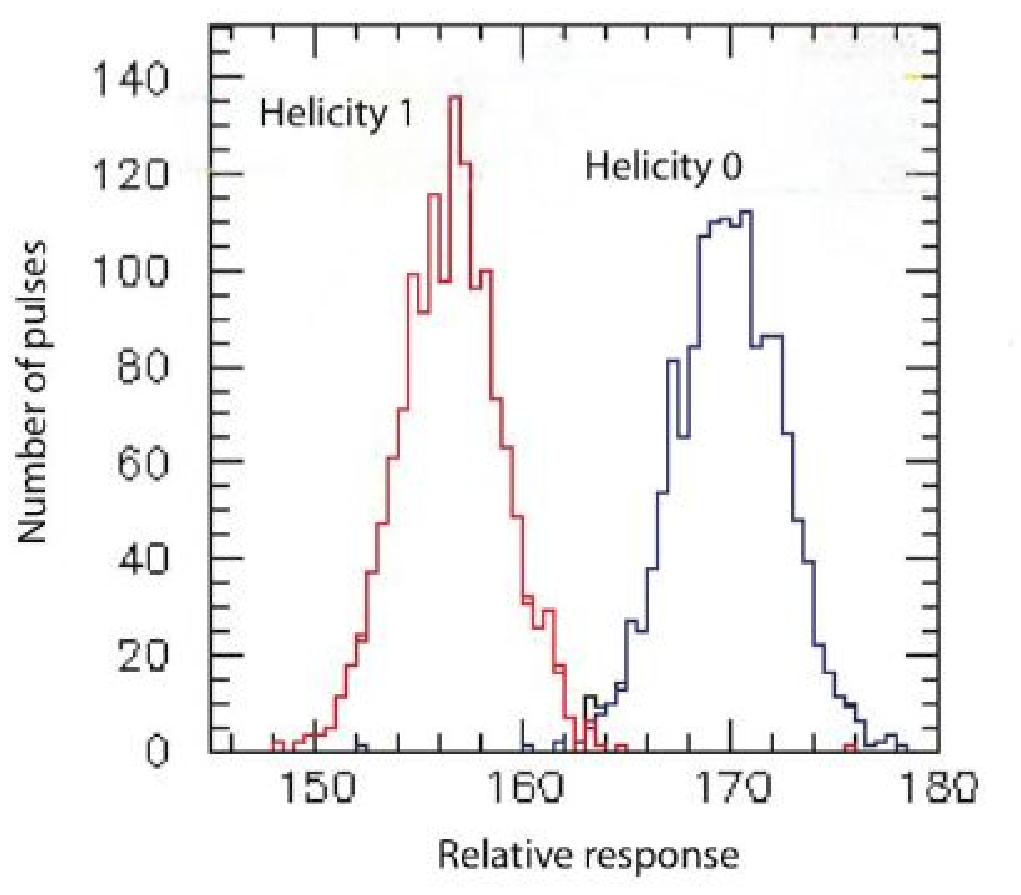,height=4in,angle=0,bbllx=-20,bblly=0,bburx=800,bbury=300,clip=,silent=}
\caption{Pulse--height  distributions  of  polarimeter  pulses  for  helicity  0  and  helicity  1 
M{\o}ller scattering off a polarized 20 $\mu$m thick iron foil and beam intensity 5.9*10$^{11}$ 
electrons per beam pulse at 10 Hz. Each polarimeter pulse was integrated over one beam pulse, and in 
this plot 
normalized according to the total
charge in the beam pulse as measured by a toroid. Note the suppressed zero on the abscissa.}
\label{fig17}
\end{figure}

Figure \ref{fig16} shows  results for two radial scans for an unpolarized carbon target. One
scan was made with the mirror set for transmission, the other for the mirror obstructing
Cherenkov photons. This lack of a sizeable signal in the obstructed case confirms that
background due to radiation interacting directly in the PMT was insignificant.
Figure \ref{fig17} shows the polarimeter response distribution obtained using a polarized foil. 
In  one  case  the  beam  polarization  is  aligned  with  the  polarization  
of target  electrons,
(total helicity 1), in the other case the polarizations are opposite (total helicity 0). The
response  is  normalized  according  to  the  charge  in  each  accelerator  beam  pulse,  as
measured by a toroidal coil.  The difference in responses arises from the dependence of
the M{\o}ller cross section on the total helicity.  The data were used to determine the extent
of the polarization of the incident electron beam. In our case this ranged from 80--90\%,
with a statistical uncertainty of about 4\% for a 20 minute run.

\section{Summary}
\label{sec4}
The scanners described in the paper turned out to be very useful in commissioning the
experiment and controlling properties of scattered beam and backgrounds in 
the E158 detector system in which electron flux scattered from the beam carrying 
$\sim$ 0.5MW of power was measured at an extremely small angle of $\sim$0.3$^\circ$. 
The scanners  were  highly radiaton resistant and provided 
very good spatial resolution of the flux profile. Knowledge of precise radial and azimuthal 
flux distributions was critical for good understanding of the detector system and minimizing 
backgrounds in the main calorimeter.
\section{Acknowledgements}
\label{sec5}

This work was supported by Department of Energy contract DE--AC03--76SF00515, and by 
the Division of Nuclear Physics at the Department of Energy and the Nuclear Physics Division of 
the National Science Fundation in the United States.


\clearpage

\begin{thebibliography}{00}
\bibitem{e158}
SLAC proposal E158: ``A Precision Measurement of the Weak Mixing Angle in M{\o}ller Scattering'', 
K.S. Kumar {\em et al.} (1997),
\ead[url] {http://www.slac.stanford.edu/grp/rd/epac/Proposal/E158.pdf}
P.L. Anthony {\em et al.}, Phys. Rev. Lett., {\bf 92}, 181602 (2004).
\bibitem{Alzak}
MIRO 2, ALANOD Aluminium--Veredlung GmbH \& Co. KG, Ennepetal, Germany.\\
Alzak is a registered trademark of the Aluminum Company of America (ALCOA).
\bibitem{geant}
GEANT v. 3.21, CERN Program Library.
\bibitem{bert}
 S.  Bertolucci et al., "Scintillation in Gases Commonly Used as Cherenkov Radiators",
 DESY preprint   75/16, July 1975.
\bibitem{uni}
Uniblitz shutter. Vincent Associates, Rochester N.Y., USA.
\bibitem{emi}
Thorn EMI, model 9083A. Electron Tubes Limited, Ruislip, Middlesex, England.
\bibitem{parker}
Parker--Daedal, model 404XR.
\bibitem{annulus}
Fabricated to our specifications by Western Mass Machining, of Holyoke, Massachusetts.
\bibitem{lv}
National Instruments, Inc., Austin, TX, USA.
\bibitem{pot}
Unimeasure Inc., Corvallis, OR, USA.
\bibitem{PMT}
Hamamatsu Photonics, model R2154--02 (2 inch). Hamamatsu City, Japan.
\bibitem{bimba}
Bimba Manufacturing Company, Monee, IL, USA.
\end{thebibliography}
\end{document}